\documentclass[preprint,tightenlines,superscriptaddress,
prd,nofootinbib,showpacs]{revtex4}

\def\beq{\begin{equation}} \def\eeq{\end{equation}}
\def\bea{\begin{eqnarray}} \def\eea{\end{eqnarray}}

\def\ts#1{\textstyle{#1}}  
    
  \def\GN{G_{\rm N}} \def\me{m_{\rm e}}
\def\half{{\textstyle{1\over2}}}

\begin{document}

\title{Comparison of QG-Induced Dispersion\\
with Standard Physics Effects}

\author{Luca Bombelli}
\affiliation{Department of Physics and Astronomy\\
University of Mississippi, University, MS 38677, U.S.A.}
\affiliation{Perimeter Institute for Theoretical Physics\\
35 King Street North, Waterloo, ON, Canada N2J 2W9\\ \ }
\author{Oliver Winkler}
\affiliation{Perimeter Institute for Theoretical Physics\\
35 King Street North, Waterloo, ON, Canada N2J 2W9\\ \ }

\date{29 Mar 2004}

\begin{abstract}
One of the predictions of quantum gravity phenomenology is that, in situations where Planck-scale physics and the notion of a quantum spacetime are relevant, field propagation will be described by a modified set of laws. Descriptions of the underlying mechanism differ from model to model, but a general feature is that electromagnetic waves will have non-trivial dispersion relations. A physical phenomenon that offers the possibility of experimentally testing these ideas in the foreseeable future is the propagation of high-energy gamma rays from GRBs at cosmological distances. With the observation of non-standard dispersion relations within experimental reach, it is thus important to find out whether there are competing effects that could either mask or be mistaken for this one. In this letter, we consider possible effects from standard physics, due to electromagnetic interactions, classical as well as quantum, and coupling to classical geometry. Our results indicate that, for currently observed gamma-ray energies and estimates of cosmological parameter values, those effects are much smaller than the quantum gravity one if the latter is first-order in the energy; some corrections are comparable in magnitude to the second-order quantum gravity ones, but they have a very different energy dependence.
\end{abstract}
\maketitle

\subsection*{Introduction}

One of the heuristic predictions from quantum gravity phenomenology that has been attracting a considerable amount of attention is the possibility that fields propagating in a quantum spacetime may exhibit dispersion, due to their interaction with the fluctuating, possibly discrete quantum geometry. In general the dispersion relation in such scenarios, whether they be motivated by loop quantum gravity, spin foams, strings, or non-commutative geometry, is obtained from a modified mass shell equation of the form \cite{Ame97,Ame04}
\beq
   c^2p^2 = E^2\,[1 + \alpha\,(E/E_{\rm QG})^\beta +
   \hbox{higher order terms}]\;, \label{mom-en}
\eeq
where $\alpha$ and $\beta$ are model-dependent constants, with $\beta = 1$ or 2 considered to be the likely values. This corresponds to propagation in a medium with an effective frequency-dependent phase velocity and  index of refraction $n(\omega) = c/v_{\rm ph}(\omega) = cp/E$, which gives
\beq
   n_{\rm QG} = 1 + \half\,\alpha\,(E/E_{\rm QG})^\beta + \hbox{h.o.t.}
    = 1 + \half\,\alpha\,(\omega/\omega_{\rm QG})^\beta + \hbox{h.o.t.}
   \label{disprel}
\eeq
The effect is expected to be very small, but it may become observable in the case of gamma rays from distant GRBs, propagating over cosmological distances of billions of light years. For such photons, we can assume the high end of the energy spectrum to be somewhat higher than 1 MeV \cite{ZM}, which means that $E/E_{\rm QG}$ is slightly larger than $10^{-22}$, if we take the quantum gravity energy scale to be the Planck energy, $E_{\rm QG} = E_{\rm P} = \sqrt{\hbar c^5/G} \simeq 10^{19}$ GeV. Observing this effect would be extremely interesting, but the smallness of the estimated numbers implies that this may be experimentally possible \cite{Ame97} in the near future only if the leading order term in Eq \ref{disprel} corresponds to $\beta = 1$.

Observationally, bounds on $n-1 \sim \Delta c/c$ around $10^{-20}$, close to the $\beta = 1$ range, have already been obtained \cite{Sch, Gha} and data have already been used to set bounds on parameters for Lorentz symmetry violating models \cite{PS}. Improved techniques may soon allow us to make more general statements about first order Planck-scale effects; as the search for quantum gravity effects on photon dispersion is pushed toward smaller orders of magnitude, it is becoming increasingly important not just to improve the observational tools, but also to be able to distinguish this effect from other actual physical effects that could either mask it or be mistaken for it (``theoretical noise"). One should therefore examine systematically other possible mechanisms, such as QED vacuum effects or couplings to other forms of classical matter, quantum fields, and geometry which can produce dispersion in gamma rays.

In this letter, we will consider effects from standard physics only. Classically, there are two possibly relevant interactions, the electromagnetic and the gravitational ones. We will begin by estimating the contribution to dispersion by the plasma effect due to electrons in the galactic and intergalactic medium, treated as free particles because of the gamma-ray frequencies involved. Then we will consider the general relativistic coupling of photons to gravity by scattering off the curved geometry, both by multiple scattering from particles and local gravitating objects of any size, and by scattering off the global geometry. Finally, we look at non-trivial vacuum effects in QED as a possible source of dispersion, by gathering results from the existing literature and applying them to our situation. In each case, the first goal is to compare the deviation of $n$ from 1 produced by the effect under consideration with the quantum gravity estimate. If the two contributions turned out to be of similar orders of magnitude in the relevant energy range, we might still be able to discriminate between them if the energy dependences are different; therefore the next important point would be to compare the frequency dependences.

A few parameter values will be needed for our estimates. In addition to a reference gamma-ray energy that we will take to be $E_\gamma = 1$ MeV, corresponding to a frequency $\omega_\gamma \approx 1.5 \times 10^{21}$ rad/s, the main obervational values we will use are \cite{WMAP} the baryon density $\rho_{\rm b} \approx 4.2 \times 10^{-28}$ kg/m$^3$, equivalent to a number density $N_{\rm b} \sim 0.25$ m$^{-3}$, the total fractional 
energy density $\Omega_{\rm tot} = 0.02 \pm 0.02$, and the Hubble parameter $H_0 \approx 71$ km/s/Mpc.

\subsection*{Dispersion Due to Classical Electromagnetic Interactions}

Cosmological gamma ray photons interact electromagnetically with charged particles they encounter, mainly electrons and protons. A detailed treatment of this interaction would consider the states of these particles (bound or free) and take into account various effects. However, the particles that most affect the photons' propagation are the lighter ones, and at the gamma-ray energies we are considering, much higher than their binding energies, the electrons can be effectively treated, for our purposes, as free particles.

As part of our underlying model, we will assume that space is homogeneous, aside from localized electromagnetic or gravitational scatterers, and we will characterize these scatterers only by their average properties, reducible to the numbers listed at the end of the previous section. Photons propagating in intergalactic space will then see a medium that we treat as a uniform plasma of free electrons, which responds to an electromagnetic wave by absorbing and re-radiating energy. As an extended distribution, the electrons produce a cumulative effect which can be described, in the high-frequency approximation, by the dispersion relation \cite{Jac}
\beq
   n = 1- {NZe^2\over2\,\epsilon_0\me\,\omega^2}\;,
\eeq
where as an approximate value for the product $NZ$ of the effective atom number density and the effective atomic number we will use the baryon number density.

With those parameter values we can estimate the effect for our reference 1-MeV photon,
\beq
   n - 1 \simeq 1.8 \times 10^{-40}\;,
\eeq
a value greater than the expected quantum gravity one for $\beta = 2$. Notice however that the value is very approximate. Furthermore, the correction term has an $\omega^{-2}$ dependence, which means that in the infinite-frequency limit there is no dispersion (geometric optics approximation), contrary to the quantum gravity case, in which high-frequency photons probe the small-scale geometry better and produce a larger effect \cite{Ame97}.

\subsection*{Dispersion from Classical Gravitational Multiple Scattering}

In general relativity, one potential source of dispersion is geometrical
scattering off objects in an extended random distribution or cluster, a
``gas" of gravitating masses. Such an effect was briefly considered in the early 1970's by Peters \cite{Pet}. In Peters' treatment, the scatterer distribution is modeled by (the continuum limit of) a sum of weak-field approximations to Schwarzschild metrics centered at locations ${\bf r}_i$, represented by a scattering potential
\beq
   \phi({\bf r}) = -\GN\sum_i{m_i\over|{\bf r}-{\bf r}_i|}\;.
\eeq
When a plane wave is incident on a layer of such a ``gas", the superposition of waves diffracted by individual objects gives rise to the effective dispersion, since each single diffraction pattern is $\omega$-dependent; this is analogous to what happens with ordinary dispersion in the atmosphere, although the scattering mechanism from individual scatterers is different in that case. Peters considered scalar, electromagnetic and gravitational waves. While for the (minimally coupled) scalar and gravitational wave cases he found the dispersion relation
\beq
   n = 1 + {2\pi\,\GN\rho\over\omega^2}\;,
\eeq
in the electromagnetic case he found that, to first order in the scatterer masses $m_i$, there was {\it no\/} frequency-dependent phase shift for a plane wave after crossing the layer of scatterers. However, the approximations he used, both in modeling the situation and in treating the quantities encountered in the calculations (some of which were divergent, as one might expect from scattering from a Coulomb-type potential), lead us to believe that his results are not very conclusive, and suggest that we do not yet discard the effect. A more cautious approach may be not to rely on the validity of Peters' calculations, but to simply notice that, independently of any model, one can estimate the size of the dispersion effect on purely dimensional grounds. Since to first order $n-1$ will be proportional to $\GN\rho$, we conclude that, {\it if\/} there is a gravitational multiple scattering effect on the index of refraction, in terms of orders of magnitude it will be at most
\beq
   n - 1 \sim {\rho_{\rm matter}\,\GN\over\omega^2}\;,
\eeq
in agreement with Peters' results for the other types of waves, where this time for $\rho$ we use the average density of all gravitating matter, roughly 6 times larger than $\rho_{\rm b}$. Notice that this estimate, as can be seen in Peters' calculations \cite{Pet}, depends only on the average $\rho$, and not on details such as the masses and sizes of individual scatterers.

If we again estimate the effect for our 1-MeV photon, we get
\beq
   n - 1 \simeq 7.5 \times 10^{-80}\;;
\eeq
thus, not only do we obtain an inverse $\omega$-dependence, consistent with what we expect from a classical effect, but the magnitude of the departure of $n$ from 1 is such that it would not compete with the quantum gravity effect even for $\beta = 3$.

\subsection*{Dispersion from Scattering off the Global Curvature --- ``Wave Tails"}

In addition to the average densities used in previous sections, a homogeneous cosmological model is characterized by a global spatial geometry and expansion rate. When waves propagate in a curved spacetime, they can scatter off the global curvature, a phenomenon that is usually described in terms of the formation of tails, or non-validity of the Huygens principle, rather than in terms of a modified index of refraction. It is known, for example, that a necessary condition for the validity of the Huygens principle in 4-dimensional spacetime is that the geometry be that of an Einstein space \cite{Gol}, and that tails generically form in the propagation of fields, both near isolated objects \cite{Man} and in a cosmological setting \cite{FG}. It would be useful therefore to analyze the latter effect in more detail in terms of modified effective dispersion relations.

The recent WMAP observational data on the microwave background are consistent with, indeed can be taken to be an indication of, a vanishing overall spatial curvature, in which case the universe on ``average" is a $k = 0$ Robertson-Walker space; such spaces are conformally flat, and since the classical Maxwell equations are conformally invariant, photons cannot develop tails from their propagation in the overall geometry. A more careful analysis (motivated by the fact that in some situations cosmological tails can be strong \cite{Nol}) would take into account the actual bounds on the spatial curvature; in this paper, we will limit ourselves to a simple comment. We can obtain a bound for the radius of curvature $R$ of space using the relationship
\beq
   R = {c\over H_0}\,{1\over|1-\Omega_{\rm tot}|^{1/2}}
\eeq
between cosmological parameters and spatial geometry, and the WMAP data \cite{WMAP}. Specifically, an indicative lower bound on $R$ can be obtained from the upper limit of the error bar on $\Omega_{\rm tot}$, giving
\beq
   R \ge 3.0\times 10^4\;{\rm Mpc} = 9.3\times10^{26}\;{\rm m}\;.
\eeq
Any classical propagation effect for photons in curved spacetime that depends only on $R$ and the wavelength will give a contribution to $n-1$ which is at most of the order of $\lambda/R$, with $\lambda \approx c/(\omega/2\pi)$. In our case, $\lambda \approx 10^{-12}$ m, and we conclude that the contribution to dispersion would be at most
\beq
   n-1 \approx {2\pi c\over\omega R} \approx 10^{-39}\;.
\eeq

\subsection*{Dispersion from QED Effects}

If we take into account the fact that a better description of gamma-ray propagation consists in treating it as taking place on the background of some (homogeneous) QED state, several effects arise which can modify their dispersion relations \cite{Sho}. These effects do not show up for light propagating in the QED vacuum for flat spacetime with no other background fields, but in a cosmological setting the speed of propagation of light can be affected by a background electromagnetic field, by the overall spacetime curvature (through vacuum polarization), and by the cosmic microwave background radiation (through photon-photon interaction). (Higher-derivative gravity theories also give rise to modified, dispersive photon propagation \cite{AB}, but we will not consider those here.)

For high-frequency electromagnetic waves propagating in a weak background magnetic field, the main parameter which determines whether the effect is dispersive or not is the number \cite{TE}
\beq
   \lambda:= {\ts{3\over2}}\,{eB^2\hbar^2\omega\over \me^3c^4}
   \,\sin\theta\;.
\eeq
Here, $B$ is the magnitude of the magnetic field, assumed to be constant, and $\theta$ the angle between $\vec B$ and the direction $\hat k$ of propagation. Dispersion occurs only if $\lambda > 1$. While the magnetic field in intergalactic space is not very well known \cite{Gio}, we can take $B \approx 10^{-7}$ G as an indicative value, at least for clusters of galaxies (it is probably smaller outside, and the fact that it is not really constant will also decrease the effective value of $\lambda$ and the overall effect); as a further overestimate, let us set $\sin\theta = 1$. We then get $\lambda \approx 4.3\times 10^{-60}\,\omega$(rad/s); even for the highest frequency $\gamma$-rays, $\lambda \ll 1$, in which case the effective index of refraction (for both polarizations) is
\beq
   n - 1 = {\alpha\,C\over4\pi}
   \left({eB\hbar\over\me^2c^2}\,\sin\theta\right)^{\!2},
\eeq
where $\alpha$ is the fine structure constant and $C$ a known polarization-dependent number of order 1. This result is independent of $\omega$, so the effect is non-dispersive (with our parameter values, this contribution to $n - 1$ is of the order of $10^{-37}$).

Let us now consider photon-photon interactions in a thermal vacuum at temperature $T$, with the aim of taking into account the effect of the CMB on propagating photons. This contribution to the index of refraction has been calculated in the low-energy situation in which electron pair creation can be neglected \cite{Bar}, where it has been shown to give the non-dispersive result
\beq
   n - 1 \approx {44\pi^2\alpha^2\over 2025}\,
   (k_{\rm B}T/\me c^2)^4\;, \label{therm1}
\eeq
and in the high-energy limit ($\hbar\omega \gg \me c^2$), where one obtains the dispersive relation \cite{LPT}
\beq
   n - 1 \approx {\alpha^2\over6}\,(k_{\rm B}T/\hbar\omega)^2
   \ln^2\left({\hbar\omega\over\me c^2}\,{k_{\rm B}T\over\me 
   c^2}\right)\;, \label{therm2}
\eeq
but the resulting values are smaller than that from Eq \ref{therm1}. Although a better estimate may eventually be necessary, we will therefore take as an indicative value of the size of the effect the one in Eq \ref{therm1}, which equals $4.7 \times 10^{-43}$ at today's temperature of 2.7 K (and $6.5 \times 10^{-31}$ at the recoupling temperature 3000 K). Therefore, we have again a value that may be somewhat larger than the one from second-order quantum gravity effects but decreases with energy.

In fact, Eq \ref{therm1} can be considered as a special case of a more general result for low-energy photons in modified QED vacua \cite{LPT}, in which $T^4$ is replaced by a quantity proportional to the excess energy density of the modified vacuum with respect to the standard one. On the one hand, as the authors pointed out, one gets a criterion for identifying situations with superluminal phase velocities. On the other hand, one may use this pattern as motivation for assuming that different modified QED vacuum effects behave in similar ways, becoming dispersive at wavelengths smaller than the Compton wavelength but giving contributions to $n - 1$ that can be bounded by extrapolating the low-energy expressions. This assumption has its limitations \cite{DG}, and more work on the actual values of various effective QED contributions to $n$ in the intermediate-energy range, and their interplay \cite{Gie}, is necessary.

Finally, for high-frequency electromagnetic waves propagating in a curved spacetime, the known general results \cite{Sho} can be summarized in the curvature-dependent ``effective light cone" in momentum space given by ($c = \GN = \hbar = 1$)
\beq
   g_{ab}\,k^ak^b - {8\pi\over\me^2}\,F\left({k^m \nabla_{\!m}
   \over\me^2}\right) T_{ab}\,k^ak^b
   + {1\over\me^2}\,G\left({k^m \nabla_{\!m}\over\me^2}\right)
   C_{abcd}\, k^ak^ca^ba^d = 0\;,
\eeq
where all indices are spacetime indices, $k^m \nabla_{\!m}$ denotes a covariant derivative along the null geodesic with tangent vector $k^a$, the functions $F$ and $G$ are {\it in principle\/} known, and $a^a$ is the polarization vector. If we take a Friedmann-Robertson-Walker metric to be a good description for the purpose of calculating the effective index of refraction, then the Weyl tensor $C_{abcd}$ vanishes because the space is conformally flat, but the phase velocity depends on $\omega$ because of the non-trivial dependence of $F$ on its argument. To find an actual expression for $n(\omega)$, even if we use the approximation that $k^a$ is constant along these null geodesics, the calculation is reduced to that of $F(\me^{-2}k^m\nabla_{\!m})\cdot T_{ab}$, for which intermediate-energy results in FRW spacetime are not available, to our knowledge. Therefore, we will again use the $C_{abcd} = 0$ low-energy, non-dispersive expression \cite{Sho}
\beq
   g_{ab}\,k^ak^b - {22\,\alpha\over45\,\me^2}\,T_{ab}\,k^ak^b = 0\;,
   \qquad{\rm or}\qquad v_{\rm ph} = 1 + {11\,\alpha\over45}\,
   {\GN\hbar^2\rho\over\me^2c^4}
\eeq
(where in the last equation all constants have been restored and we have used the matter-dominated universe approximation of vanishing pressure), to bound the desired value. If we set $\rho = \rho_{\rm matter}$, our estimate then is
\beq
   |n-1| < 5 \times 10^{-82}\;;
\eeq
clearly, our conclusions regarding this effect are not very sensitive to minor improvements in the approximations used.

\subsection*{Comments}

To summarize, none of the effects we considered gives a contribution to $n - 1$ for $\gamma$-rays of energies in the MeV range that can compete with a quantum gravity effect as described by Eq \ref{disprel} with $\beta = 1$, i.e., of order $10^{-22}$ in this energy range. It seems plausible that tighter observational bounds of this magnitude will be available in the not-too-distant future \cite{SUV}, and we should therefore consider looking for second-order effects in $E/E_{\rm P}$.

A few possible sources of dispersion have not been included in our discussion so far. An obvious one is a possible photon mass. Most current bounds on $m_\gamma$ are around $10^{-16}$ eV \cite{Jac}, although some much tighter bounds exist from the galactic magnetic field; even with that value, and treating the photon like a relativistic massive particle, a 1-MeV photon has $n - 1 \approx m_\gamma^2c^4/2E^2 < 10^{-44}$, a bound similar to (but somewhat smaller than) others we obtained above that have an inverse $\omega$ dependence and may compete in magnitude with $\beta = 2$ quantum gravity effects. Another interesting effect, of a somewhat different kind, may arise from the possible multi-valued nature of dispersion relations of the type (\ref{mom-en}), which could manifest itself in birefringence \cite{Leh}.

A less obvious (and less easy to evaluate) additional mechanism is related to the presence of inhomogeneities in the universe. Using cosmological models characterized just by the average values of physical quantities is appropriate for many purposes, but is not always a good approximation. In general, qualitatively new phenomena may appear when considering local fluctuations (see, e.g, Ref \cite{HS}); for example, our argument concerning multiple gravitational scattering breaks down as soon as we consider scatterers of finite size and mass, which provide additional dimensional parameters, and a general feature of local inhomogeneities is that their effect may cancel out {\it on average\/} but not as far as fluctuations are concerned. There is probably no immediate need to obtain results for these effects, but while ideas and techniques are being developed to look for second-order quantum gravity effects, one should also look into ways of tightening the bounds we listed above, and filling in our omissions.

\subsection*{Acknowledgements}

This work was supported in part by NSF grant number PHY-0010061 to the
University of Mississippi. We are grateful to H Gies, R Lehnert, and S Sarkar for helpful comments, and LB would like to thank Perimeter Institute for hospitality.

\end{document}